\documentclass[%
 reprint,
 amsmath,amssymb,
 aps,
]{revtex4-1}
\usepackage{amsmath}
\usepackage{amssymb}

\usepackage{graphicx}
\usepackage{color}

\graphicspath{{images/}}

\newcommand{\be}{\begin{equation}}
\newcommand{\ee}{\end{equation}}
\def\ba{\begin{aligned}}
\def\ea{\end{aligned}}

\newcommand{\bea}{\begin{eqnarray}}
\newcommand{\eea}{\end{eqnarray}}

\newcommand{\Tr}{{\rm \, Tr\,}}

\renewcommand{\Im}{{\rm \, Im\,}}

\renewcommand{\hat}[1]{{\widehat #1}}

\renewcommand{\Im}{{\rm Im\,}}

\begin{document}

\title{Quantum transport through the edge states of Zigzag phosphorene nanoribbons  in presence of a single point defect: analytic Green's function method}
\author{M. Amini}
\email{msn.amini@sci.ui.ac.ir}
\author{M. Soltani}  
\affiliation{Department of Physics, University of Isfahan(UI)—Hezar Jerib, 81746-73441, Isfahan, Iran} 
\begin{abstract}
Zigzag phosphorene nanoribbons have quasi-flat band edge modes entirely detached from the bulk states.
We analytically study the electronic transport through such edge states in the presence of a localized defect for semi-infinite and finite ribbons.
Using the tight-binding model, we derive analytical expressions for the Green's function and transmission amplitude 
of both pristine and defective nanoribbons. We
find that the transmission of both semi-infinite and finite ribbons is 
sensitive to the location of a single impurity defect with respect to the edge.
By the presence of an impurity on the outermost edge site of the ribbon, the transmission through the edge channel,  similar to a one-dimensional chain, strongly suppresses for the entire energy spectrum of the quasi-flat band.
In contrast, the transmission of  low-energy $(E\approx 0)$ states, is robust as the impurity is moved  one position far  away  from  the  edge on the same sub-lattice.
The analytical calculations are also complemented by exact numerical transport computations using the Landauer approach.

\end{abstract}


\maketitle 

\section{Introduction}

A  single  layer  of  black  phosphorus(BP) in which P atoms arranged in a hexagonal staggered lattice called phosphorene.
This promising new two-dimensional (2D) material, in the sense of its applications in
nano-electronics, can be exfoliated from bulk BP due to weak interlayer
van der Waals interaction~\cite{Li2014,Liu2014,Neto2014}. 
A striking structure property of phosphorene is the bonding of each P atom with other three nearest P atoms via $sp^3$
hybridization which leads to forming a puckered honeycomb lattice exhibiting radically  different
anisotropic properties in electronic, mechanical, thermal, and transport quantities \cite{Neto_PRL2014,Qiao2014,Peeters2014,Guinea2014,Yang2014,Katsnelson2015,Asgari2015}.

Of late, 2D nanomaterials have attracted  lots of interest and phosphorene, by patterning into phosphorene
nanoribbons (PNRs) which can be fabricated with lithography and plasma etching of BP, is a new candidate material for novel electronic devices\cite{Feng2015,Maity2016,Fang2017}. 
At the same time, there are two typical types of PNRs namely armchair PNRs (aPNRs) and zigzag PNRs (zPNRs) depending on the directions that phosphorene is cut along.
However, different PNRs shows different physical properties.
From the band structure point of view, pristine aPNRs are semiconducting, while pristine zPNRs are metallic \cite{Carvalho2014}.
Furthermore, due to the large out-of plane hopping parameter in  zPNRs,  there exsist two quasi-flat edge modes which are completely isolated from the bulk  bands\cite{Ezawa2014} while PNRs don't have  such edge states\cite{Asgari2017}.

Similarly to other materials, structural defects is practically an inevitable factor in real PNRs.
At the same time, artificial defects may also use to design and functionalize the materials for new applications.
For the case of phosphorene and bulk BP, first principle calculations showed\cite{vacancy-DFT2015} that  different types of point vacancy defects can be formed in phosphorene which results in different electronic band structures.
It is also shown that atomic vacancies in phosphorene are highly itinerant at low temperatures\cite{vacancy-DFT2017}. 
Furthermore, utilizing a combination of low-temperature scanning tunneling microscopy/spectroscopy  and electronic structure calculations, single atomic vacancies observed on the surface of bulk BP\cite{vacancy-exp-2017}.  
On the other hand, very few studies\cite{Wu2017,Poljak,Peeters2018,Asgari2018} are currently available on the defective PNRs.
For instance, the electronic transport properties of defective PNRs containing atomic vacancies is studied recently\cite{Peeters2018}. 
Also, it is found that single vacancies can create quasi-localized states, which can affect the conductance of PNRs and 
the effect of doping on the charge transport of zPNRs studied in \cite{Asgari2018}.

It is however, an intersting question to understand the quntum transport in a quasi-flat band composed of edge states in presence of a localized impurity potential. 
In this paper we study the effect of a single impurity on the quantum transport of the zPNRs 
focusing on the dependence of transmittance on the position of impurity. 
We consider zPNRs
with different widths and calculate the transmittance through the edge states both analytically and numerically.  
The paper is organized as follows:
In section \ref{II} we generally
introduce our model and formalism of calculating and computing the 
the transmission coefficient of zPNRs using the tight-binding Hamiltoninan with the scattering approach. 
Section \ref{III} is devoted
to discussion on the results obtained for different impurity positions of both semi-infinite and finite zPNRs.
Finally, we wrap up the paper with the conclusion in section \ref{IV}.

\begin{figure}[t!]
	\center{\includegraphics[width=1.1\linewidth]{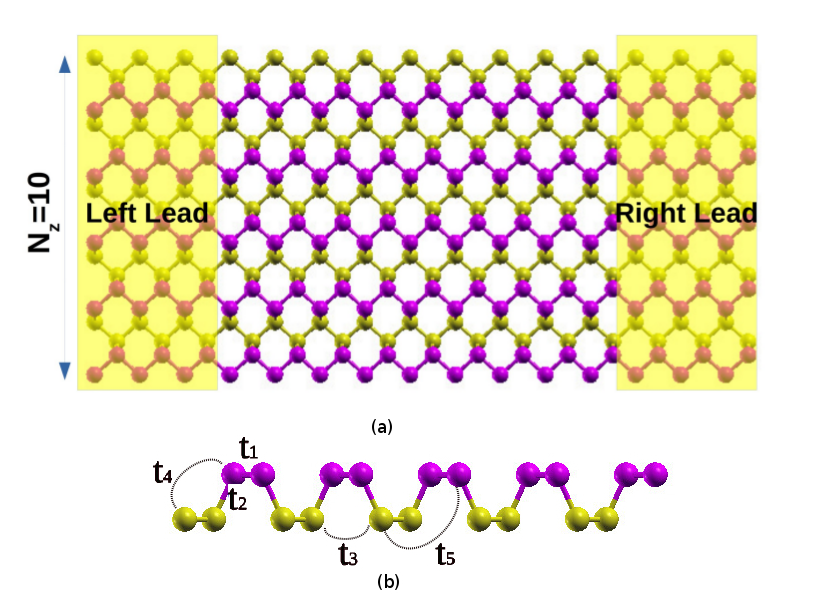}} 
	\caption{(Color online) (a) Schematic representation of  a zPNR  with $N_z=10$ zigzag chain across the width of ribbon.
	The shadow areas represent the source (left) and drain (right) semi-infinite leads.
	(b) Hopping parameters used to describe the tight-binding Hamiltonian of the zPNR.
	}
	\label{fig1}
\end{figure}
\section{MODEL and FORMALISM}\label{II}


The atomic structure of the pristine zPNR used in our study
is  shown  in  Fig.~\ref{fig1}(a).
In the tight-binding approximation, this structure can be well represented\cite{Ezawa2014,Rudenko} by the following Hamiltonian:
\be
H=H_0 + H_1= \sum_{i,j} t_{ij} c^\dagger_{i} c_{j} + \sum_{i} \varepsilon_{i} c^\dagger_{i} c_{i},
\label{HAM}
\ee
where  $t_{ij}$ is the hopping integral between sites $i$ and $j$, $\varepsilon_i$ is the on-site
energy at site $i$ describing impurity (defect) scattering, and $c^\dagger_i (c_i )$ is the creation (annihilation) operator of an electron
at site $i$. In this model, five hopping integrals which is shown in Fig.~\ref{fig1}(b) will be considered as 
$t_1 = -1.220$eV, $t_2 = 3.665$eV, $t_3 = -0.205$eV, $t_4 = -0.105$eV, and $t_5 = -0.055$eV\cite{Rudenko}. 
In the absence of defects, in our calculation, the on-site energies $\varepsilon_{i}$ are
set to zero for all lattice sites and will be non-zero for the certain defective sites in the defective system.
From now on, we will express all energies in the units eV.
The transmission calculation of defective semi-infinite and finite zPNRs is performed analytically by solving  
the scattering problem for an electron moving on the zPNR lattice with localized impurity.
We will obtain an analytic solution for this system assuming that the electron has initially momentum $k$ in the left-hand side of the zPNR.
In presence of the defects, the electron will be scattered by the defects and will be partially reflected and partially transmitted.
The reflection and transmission amplitudes $\mathcal{R}$ and $\mathcal{T}$ can be studied using the so-called $\hat{T}$ matrix approach\cite{Economou}.
This analytical approach is described in detail in the following section.

To be able to check our analytic solution, we also use the  Landauer approach that is widely used to study quantum transport properties at equilibrium and works on the basis of the recursive Green’s function technique\cite{Datta}.
In this method, we consider a two terminal device consisting of left lead, scattering region, and right lead (Fig.~\ref{fig1}(a)).
Then, the transmission coefficient (transmittance) $T=|\mathcal{T}|^2$ from the left lead to the right lead   can be conveniently written as\cite{Caroli} $T(E)=\Tr{[\Gamma_L G(E) \Gamma_R G^\dagger(E)]}$
which relates the transmittance
$T(E)$ at a given carrier energy $E$ to the Green’s function $G(E)=(E-H_{\text{center}}-\Sigma_L(E)-\Sigma_R(E))$ and line width function $\Gamma$.
The line width function including the coupling
between transport channel with source and drain leads can be calculated as $\Gamma_{L(R)}(E)= i [\Sigma_{L(R)}(E)-\Sigma_{L(R)}^\dagger(E)]$ where $\Sigma_{L},\Sigma_{R}$ are the self-energies.
Here $H_{\text{center}}$ denotes the Hamiltonian of the scattering region including the effect of the impurity potential.

\subsection{SEMI-INFINITE PHOSPHORENE}
The electronic properties of PNRs
are mainly affected by the quantization effect across the width of the ribbon due to the appearance of edges.
To proceed, it is important to consider two different situations, namely the case of zPNR with a single edge represented by a semi-infinite zPNR and the case of finite ribbon with two parallel edges.
In this section, we present and discuss the quantum transport of the semi-infinite zPNR case and 
the case of finite zPNR will be considered in the next section.
The analytical calculation thus can be performed considering a semi-infinite plane with a
single edge ($N_z\rightarrow\infty$).
On the other hand side, for the numeric computations, we assume the ribbon to be wide enough such that one of the edges can safely be ignored~\cite{Fazileh}. The latter can be achieved by checking the sensitivity of edge transmission to system size practically.

\begin{figure}[t!]
\center{\includegraphics[width=1\linewidth]{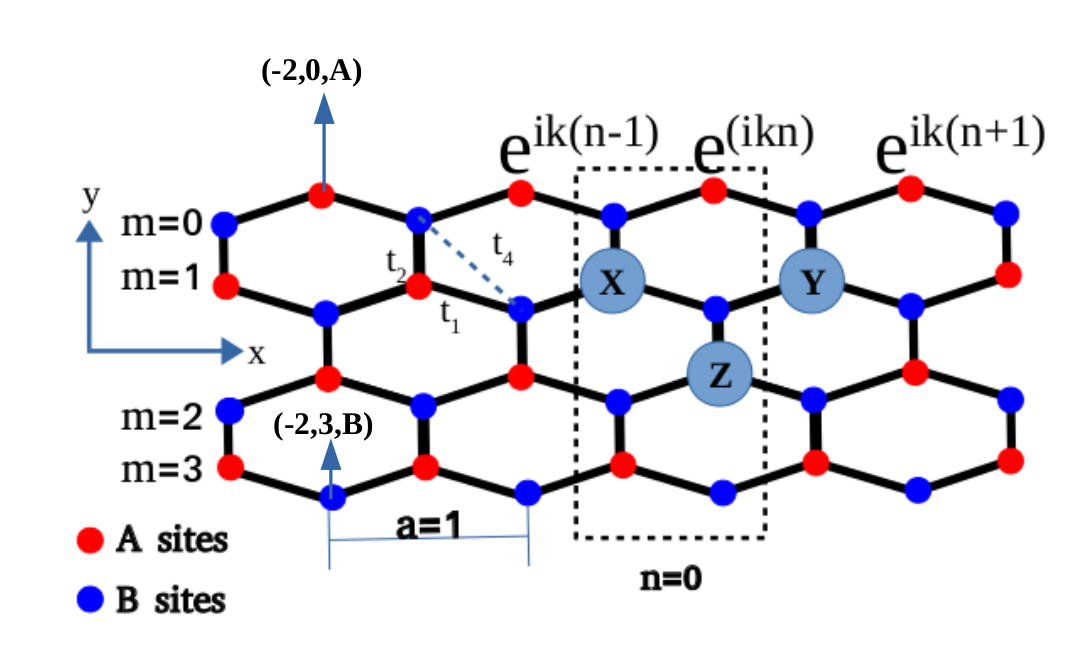}} 
\caption{(Color online) Schematic representation of the  anisotropic honeycomb-lattice model with a
zigzag edge in the $y$ direction which is used to describe zPNRs. The components of wave-function  on the edge sites indicated by $..., e^{ik(n-1)} , e^{ikn} , e^{ik(n+1)} , ...$ with  site location $n$ on the edge and wave-number $k$ which is normalized
by the length scale $a$ of the primitive translation vector of each zPNR. Each lattice site belonging to the $n$th supercell (which is shown by dashed rectangle for $n=0$) and $m$th zigzag chain from the edge can be described by $(n,m,\nu)$ with $\nu = A, B$ referring to the sub-lattices.}
\label{fig2}
\end{figure}

\subsubsection{Clean semi-infinite phosphorene}
Let us start with the perfect system  where there is no impurity defect. 
The clean phosphorene confined system with zigzag boundary supports an interesting edge states around $E_F$.
These edge states of zPNRs result in a quasi-flat band separated totally from the bulk bands\cite{Ezawa2014}.
For convenience, we can assume phosphorene as an anisotropic honeycomb-lattice model\cite{Ezawa2014} with the translational symmetry in the $x$ direction which is shown in Fig.~\ref{fig2}. This lattice consists of two sub-lattices "A" and "B" which is shown with "red" and "blue" sites respectively in  Fig.~\ref{fig2}.
Here, each lattice site belonging to the $n$th supercell and $m$th zigzag chain from the edge can be described by $(n,m,\nu)$ with $\nu = A, B$ referring to the sub-lattices.
From now on, we choose the lattice constant $a$ which is shown in Fig.~\ref{fig2} as the length scale and consider $a=1$.

Using the anisotropic honeycomb-lattice model, we can rewrite the Hamiltonian (\ref{HAM}) in terms of this new sub-lattice creation and annihilation operators $a^\dagger_{n,m},a_{n,m},b^\dagger_{n,m},b_{n,m}$ and since the hopping integrals $t_1$ and $t_2$ are much larger than the others, we can set $t_3=t_5=0$ for the analytic solution.
Later, we will take into account the effect of $t_4$ perturbatively.
Therefore, we can write
\bea
&H_0=&H'+H'', \nonumber \\
&H'=&\sum_{n,m} t_1 (a^\dagger_{n,m}+a^\dagger_{n+1,m}) b_{n,m} + t_2 a^\dagger_{n,m} b_{n,m+1} +H.C., \nonumber\\
&H''=&\sum_{n,m} t_4(a^\dagger_{n,m+1}+a^\dagger_{n+1,m+1})a_{n,m}+H.C., \nonumber\\
&&+\sum_{n,m} t_4(b^\dagger_{n,m+1}+b^\dagger_{n-1,m+1})b_{n,m}+H.C.
\label{newH}
\eea

In the absence of $H''$ in the Hamiltonian(\ref{newH}), which breaks the particle-hole symmetry, the resulting edge states of the system form a perfectly flat band in the middle of the energy spectrum\cite{Ezawa2014}.
We can start constructing the analytic wave-function
for these zero energy modes by labeling the wave-function components  on the edge sites  of zPNR with $..., e^{ik(n-1)} , e^{ikn} , e^{ik(n+1)} , ...$ (Fig.~\ref{fig2}) where 
the wave-number $k$ is normalized
by the length scale $a$. 
Therefore, the corresponding eigenvalue problem for the edge state, $H'|\Psi\rangle=0$, implies
vanishing of the total sum of the components of the complex wave function over the nearest-neighbor sites. 
This implies the living of wave function entirely on only one of the sub-lattices namely "A" sub-lattice in Fig.~\ref{fig2}. So, we can  label this wave-function as $|\Psi^A\rangle$. 
Now, considering the other components of this wave-function $|\Psi^A\rangle$ at each  neighboring site of the sub-lattice as  $X$, $Y$, and $Z$ which is shown in Fig.~\ref{fig2}, we have the following set of equations:
\bea
&&t_1(e^{ik(n-1)}+e^{ikn})=-t_2X\nonumber \\
&&t_1(e^{ikn}+e^{ik(n+1)})=-t_2Y  \nonumber \\  
&&t_1(A+B)=-t_2Z,
\label{Eq2}
\eea
which can be solved as:
\bea
&& X=-2\frac{t_1}{t_2}\cos{(\frac{k}{2})}e^{ik(n-\frac{1}{2})}\nonumber \\
&& Y=-2\frac{t_1}{t_2}\cos{(\frac{k}{2})}e^{ik(n+\frac{1}{2})}  \nonumber \\  
&& Z=(-2\frac{t_1}{t_2}\cos{(\frac{k}{2})})^2e^{ikn}.
\label{Eq3}
\eea
We now apply this argument again, with $m=0$ zigzag chain replaced by $m=1$ to obtain the same form of the solution.
Thus, the wave-function components at each
site of the $m$th zigzag chain from the edge, $\Psi^A_m$, is proportional to $\alpha^m(k)$ where $\alpha(k)=-2\frac{t_1}{t_2}\cos{(\frac{k}{2})}$ and clearly shows the localization of the edge state wave-function on the edge sites. 
This means that 
\be
|\Psi^A(k)\rangle=\frac{1}{\sqrt{\pi}}\sum_{n,m} \alpha^m(k) \gamma(k)e^{ik(x_A)}|n,m,A\rangle
\label{Psi_k}
\ee
describes an edge state localized on $A$ sub-lattice near the edge $m=0$ with zero amplitudes on $(n,m,B)$ sites.  
Here, the $x_A$ coordinate is chosen in such a way that $x_A(n,m) = n$ if the site $(n,m,A)$ sits to the right of supercell (even $m$) and otherwise, $x_A(n,m)=n-\frac{1}{2}$. 
The coefficient $\gamma(k)$ is the normalization factor of the wave-function and it is straightforward to show $\gamma^2(k)=1-\alpha^2(k)$.

In the remainder of this subsection, we focus on
the  calculation of the effect of hopping parameter $t_4$ on the flat band which we obtained till now. 
Using the above wave-function and calculating the expectation value of $H_1$ term by standard first-order perturbation theory, one can easily obtain the following energy spectrum:
\begin{align}
E(k)=\langle \Psi^A|H''|\Psi^A \rangle &=-4\frac{t_4t_1}{t_2}(1+\cos{(k)}) \nonumber \\
&=\varepsilon_0-2t'\cos{(k)}.
\label{Eq4}
\end{align}
where $\varepsilon_0=-4\frac{t_4t_1}{t_2}$ is an energy shift.
This resembles the energy spectrum of a one-dimensional (1D) chain with hopping parameter $t'=2\frac{t_4t_1}{t_2}=0.07eV$ to the nearest neighbor sites.
It is evident that up to this first-order perturbation calculation, the wave-function will not change since the bulk states are separated from the edge one with a large energy gap which leads to zero correction to the first order perturbation.

\begin{figure}[t!]
\center{\includegraphics[width=1\linewidth]{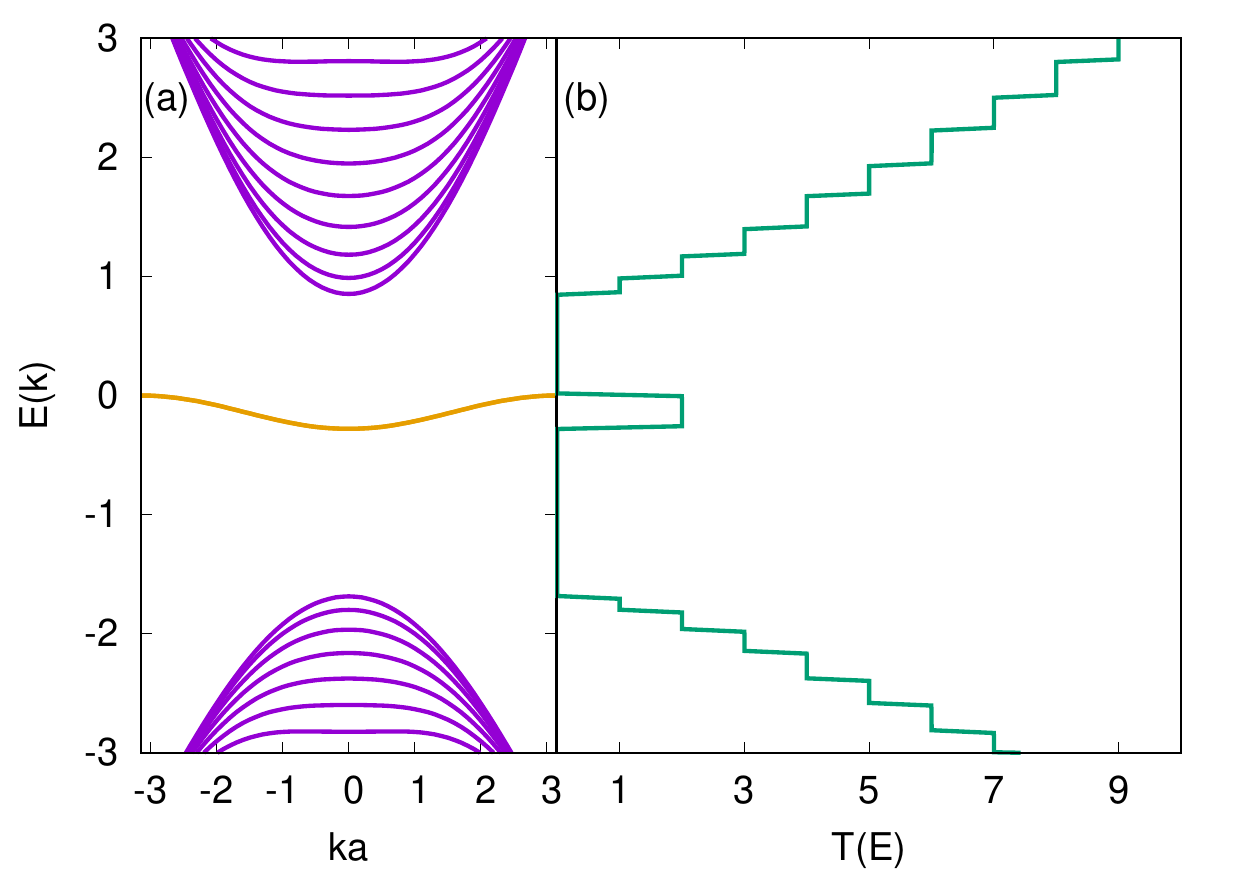}} 
\center{\includegraphics[width=1\linewidth]{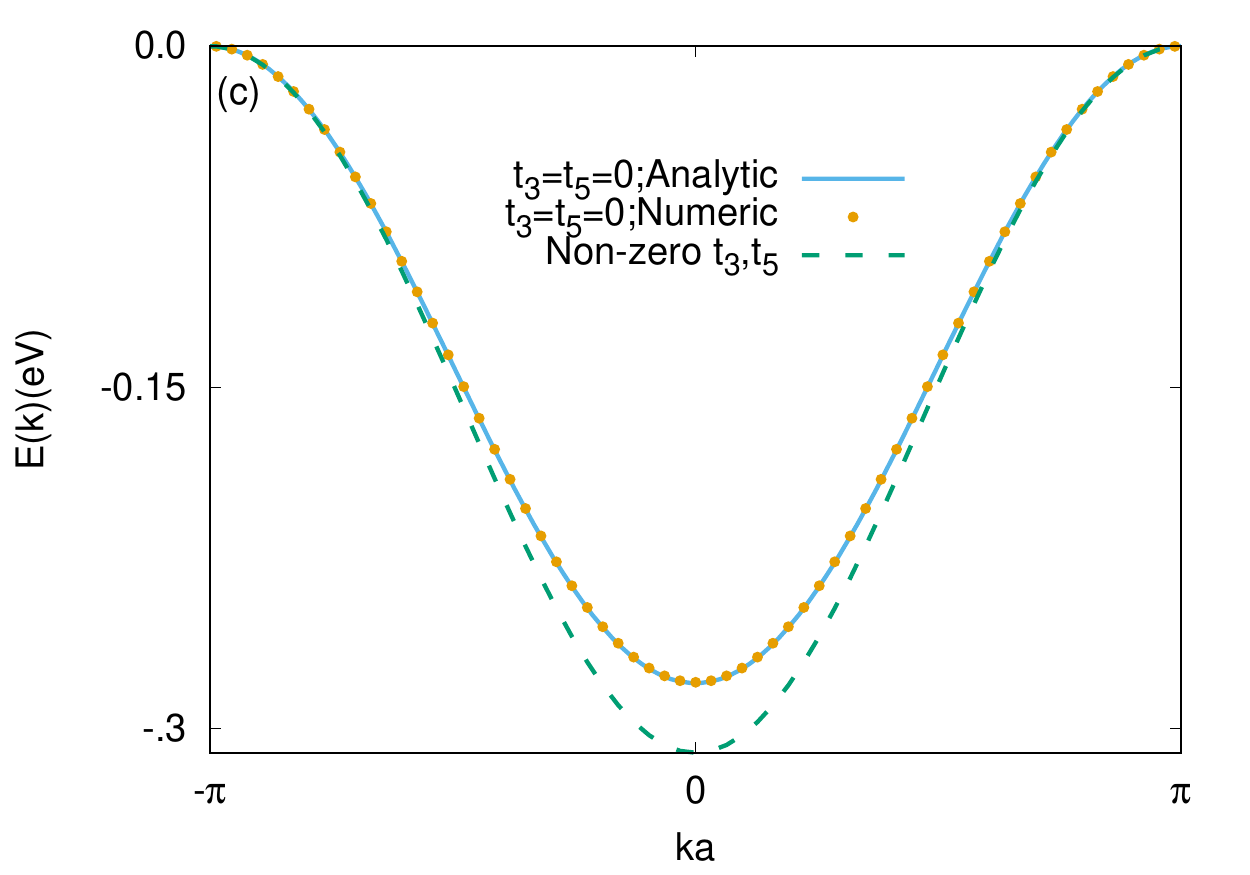}} 
\caption{(Color online) Numerical electronic band structure of a clean zPNR with $N_z=30$ (a) and the transmittance of the same system (b). A comparison between the edge states energy dispersion which is obtained analytically with the one obtained numerically (orange points) both in presence  (green dashed line) and absence  (blue solid line) of $t_3$ and $t_5$ (c).}
\label{band-clean1}
\end{figure}

Fig.~\ref{band-clean1} (a) shows the low-energy spectrum of the clean zPNR of width $N_z=30$.
The result is obtained by numerically solving the Hamiltonian (\ref{HAM}) in the absence of $t_3$ and $t_5$.
The corresponding transmittance of the system is also shown in Fig.~\ref{band-clean1} (b).
It is obvious that the two-fold degenerate quasi-flat band of the edge states is entirely separated from the bulk bands in the middle of the gap in Fig.~\ref{band-clean1} (a). 
This results in the appearance of two transmission channel in the gap region which is presented in Fig.~\ref{band-clean1} (b).
Fig.~\ref{band-clean1} (c) shows detail comparison between numerical edge band with the one obtained analytically in Eq.(\ref{Eq4}).
The numerical results (orange points) and the analytical expression of Eq.(\ref{Eq4}) (blue solid line) show an excelent agreement for $t_3=t_5=0$. 
The presence of $t_3$ and $t_5$ (green dashed line) makes no big qualitative difference with respect to the analytical expression of Eq.(\ref{Eq4}) which is a necessary ingredient in calculating the lattice Green’s function later.  

\subsubsection{Lattice Green’s function of zPNR}
The aim of this subsection is to obtain an exact expression for the lattice Green's function of the semi-infinite zPNR.
We are interested in the transport through the edges of zPNR so, we use the wave-function in Eq. (~\ref{Psi_k})
which is obtained for a defect-free zPNR. 
The associated retarded Green’s function is then given by
\begin{align}
&G_0(n,m,\nu;n',m',\nu';\varepsilon)= 
\int_{-\pi}^{\pi} dk \frac{|\Psi^\nu(k)\rangle \langle\Psi^\nu(k)|}{\varepsilon-E(k)+i0^+}
\nonumber \\
&=\int_{-\pi}^{\pi} \frac{dk}{2\pi} \frac{e^{ik[x_\nu(n,m)-x_\nu'(n',m')]}\alpha^{m+m'}(k)\gamma^2(k)}{\varepsilon-\varepsilon_0-2t'\cos{(k)}+i0^+},
\end{align}
where the integral is over the $1$-st Brillouin zone, defined by $-\pi<k<\pi$. 

The typical integrals appearing in the above equation
can be evaluated by means of the residue theorem.
To be able to understand, we first consider the diagonal element $m=m'=n=n'=0$ and $\nu=\nu'=A$.
Therefore, we need to evaluate the following integral
\be
G_0(0,0,A;0,0,A;\varepsilon)= \frac{1}{2\pi}\int_{-\pi}^{\pi} dk \frac{1-\alpha^2(k)}{\varepsilon-\varepsilon_0-2t'\cos{(k)}+i0^+},
\ee
which can be separated into two terms as
\be
G_0(0,0,A;0,0,A;\varepsilon)= \left[1-2(\frac{t_1}{t_2})^2\right]I_1-(\frac{t_1}{t_2})^2(I_2+I_3)
\label{Iss}
\ee
where
\begin{align} 
I_1=\frac{1}{2\pi}\int_{-\pi}^{\pi} dk \frac{1}{\varepsilon-\varepsilon_0-2t'\cos{k}+i0^+}, \\ \nonumber
I_2=\frac{1}{2\pi}\int_{-\pi}^{\pi} dk \frac{e^{ik}}{\varepsilon-\varepsilon_0-2t'\cos{k}+i0^+}, \\ \nonumber
I_3=\frac{1}{2\pi}\int_{-\pi}^{\pi} dk \frac{e^{-ik}}{\varepsilon-\varepsilon_0-2t'\cos{k}+i0^+}. 
\end{align} 

\begin{figure}[t!]
\center{\includegraphics[width=.45\linewidth]{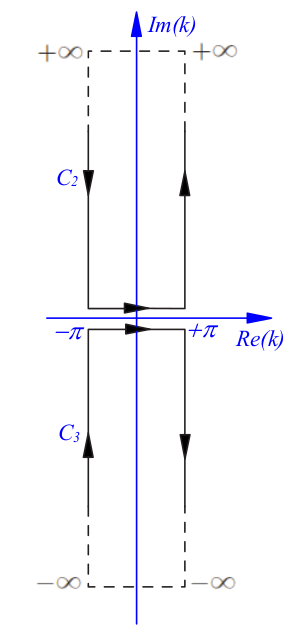}} 
\caption{(Color online) Schematic depiction of integration countour for integral $I_2$ and $I_3$. The countour $C_2$ is used for $I_2$ and $C_3$ for $I_3$.}
\label{countour-fig}
\end{figure}
The integral $I_1$ can be evaluated simply by transforming it into an integral over the complex variable $z=e^{ik}$ to set $\cos{(k)}=(z+z^{-1})/2$ and $dk=idz/z$.
By doing so, the denominator appears in the form of a
second-order expression in terms of $z$ with two solutions.
These solutions are the simple poles of the integrand. By
closing the integration contour with a unit radius circle
around the origin of the complex plane, only one of the
poles occurs inside the contour. Employing the residue
theorem, the derivation results in
\be
I_1=\frac{-i}{2t' \sin{(k_0)}}, 
\ee
where
\be
k_0=\cos^{-1}{\left[\frac{\varepsilon-\varepsilon_0}{2t'}\right]}.
\label{E_def}
\ee
By the same token, it is easy to show that the integrand in $I_2$ has two simple poles at
$k_1=k_0+i0^+$ and $k_2=-k_0-i0^+$ with the same residues of 
$\frac{e^{+ik_0}}{2t' \sin{(k_0)}}$.
In this case, performing the integral we complete the integration contour
by an infinite rectangle in the upper half-plane as shown
in Fig.~\ref{countour-fig}.
The integrand over this contour vanishes as $\Im{(k)}\rightarrow+\infty$.
On the other hand, due to the periodicity of the integrand the vertical paths have no contribution to the integral.
Therefore, since the point $k_1$ is within and $k_2$ is exterior to the
contour, the final result for $I_2$ is
\be
I_2=\frac{e^{+ik_0}}{2t' \sin{(k_0)}}.
\ee

Correspondingly, one can also obtain the analytic solution of integral $I_3$.
Doing so, since the exponential in the integrand is negative, the contour should be
completed by an infinite rectangle in the lower half-plane as shown in Fig.~\ref{countour-fig}.
In this case, only $k_2$ occurs inside the contour. This means that the resulting expression for $I_3$  is as expected equal to
what we obtained for $I_2$.

It is now easy to substitute the closed-form expressions for $I_1$, $I_2$ and $I_3$ into Eq.(\ref{Iss}) which results in
\be
G_0(0,0,A;0,0,A;\varepsilon)=\frac{-i\gamma^2(k_0)}{2t'\sin{(k_0)}}-\frac{(\frac{2t_1}{t_2})^2}{4t'}.
\label{G0A_D}
\ee

Following the same analysis, it is straightforward to
obtain the analytical form of the off-diagonal element
$G_0(n,0,A;0,0,A;\varepsilon)$ as
\be
G_0(n,0,A;0,0,A;\varepsilon)=\frac{-i\gamma^2(k_0)e^{ik_0n}}{2t'\sin{(k_0)}}.
\label{G0A_OD}
\ee
We need also to calculate the matrix elements $G_0(0,1,A;0,1,A;\varepsilon)$ and $G_0(n,1,A;0,1,A;\varepsilon)$ later. To do so, we use the same method which results in 
\begin{align}
G_0(0,1,A;0,1,A;\varepsilon)=&\frac{1}{2\pi}\int_{-\pi}^{\pi}dk\frac{\gamma^2(k)\alpha^2(k)}{\varepsilon-\varepsilon_0-2t'\cos{(k)}+i0^+}& \nonumber \\ 
=\frac{-i\gamma^2(k_0)\alpha^2(k_0)}{2t'\sin{(k_0)}}
+&\frac{(\frac{2t_1}{t_2})^2(1-(\frac{2t_1}{t_2})^2)}{4t'}-\frac{(\frac{2t_1}{t_2})^4\cos{(k_0)}}{8t'},&
\label{G0B_D}
\end{align}
and similarly
\be
G_0(n,1,A;0,1,A;\varepsilon)=\frac{-i\gamma^2(k_0)\alpha^2(k_0)e^{ik_0n}}{2t'\sin{(k_0)}}.
\label{G0B_OD}
\ee
These results will be used in the following sections to obtain the transmition amplitiude and LDoS for defective zPNRs.

\subsubsection{Semi-infinite phosphorene with a single defect}
Let us now consider semi-infinite phosphorene in the presence of a point defect. 
Suppose we have an impurity on site $(n_0,m_0,\nu_0)$ which can be described by the following Hamiltonian:
\be
H_1=U|n_0,m_0,\nu_0\rangle \langle n_0,m_0,\nu_0|,
\label{defect_H}
\ee
where $U$ denotes the on-site potential of the defected site.
The effect of this single defect on the electronic states can be studied using the so-called transition matrix ($\hat{T}$) approach\cite{Economou}. It is formally defined as
$\hat{T}=\frac{H_1}{1-H_1G_0}$ where $G_0=(\varepsilon+i\eta-H_0)^{(-1)}$ is the defect-free Green's function which we introduced before. 
Within $\hat{T}$-matrix approach, the corresponding wave-function of the defective system, $|\Psi\rangle$, can be obtained from the defect-free state, $|\Psi_0\rangle$, 
by the Lippmann-Schwinger equation\cite{Economou}
\be
|\Psi\rangle=|\Psi_0\rangle+\hat{G_0}\hat{T}|\Psi_0\rangle. 
\label{LS}
\ee
From an expression of this form, it is easily possible to read the transmition ampilitude $\mathcal{T}$ of an electron with momentum $k_0$ as 
\be
\mathcal{T}=1+\frac{UG_0(n,m_0,n_0,m_0,\varepsilon)e^{-ik_0n}}{1-UG_0(n_0,m_0,n_0,m_0,\varepsilon)},
\label{T_def}
\ee
where $n>0$.

Furthermore, the Green’s function of the defective system, $\hat{G}$,  can be expressed in terms of the clean (defect-free) Green’s function, $\hat{G_0}$, according to Dyson’s equation
\be
\hat{G}=\hat{G_0}+\hat{G_0}\hat{H_1}\hat{G}.
\ee
Using this expression, we can write the local density of states (LDoS), $\rho(n,m,\nu;\varepsilon)$, in the site of the position $(n,m,\nu)$ directly:
\be
\rho(n,m,\nu;\varepsilon)=-\frac{1}{\pi} \Im{G(n,m,\nu;n,m,\nu;\varepsilon)}.
\label{rho-def}
\ee
It is straightforward to show that
\begin{align}
&\rho(n,m,\nu;\varepsilon)=\rho_0(n,m,\nu;\varepsilon)\nonumber\\
&-\frac{\Im{}}{\pi}\{\frac{UG_0(n,m,\nu;n_0,m_0,\nu_0;\varepsilon)G_0(n_0,m_0,\nu_0;n,m,\nu;\varepsilon)}{1-UG_0(n_0,m_0,\nu_0;n_0,m_0,\nu_0;\varepsilon)}\}.
\label{LDOS-def}
\end{align}
These expressions will be used in the next section to study the transport through edge states of zPNRs in presence of a single defect.
\subsection{FINITE zPNR}

In this section, we study the quantum transport of a finite width zPNR in the presence of a single defect.
As we already mentioned in the preceding sections, the wave function $|\Psi^A\rangle$ (a state with nonzero amplitude only on "A" sub-lattice) satisfies the eigenvalue equation $H'|\Psi^A\rangle=0$ for a wide zPNR. 
However, this equation is no longer valid for a finite width ribbon and it is a simple matter to obtain the following off-diagonal matrix elements:
\begin{align}
&\langle \Psi^B(k)|H'|\Psi^A(k)\rangle &=\langle \Psi^A(k)|H'|\Psi^B(k)\rangle \nonumber \\
&&= t_2 \gamma^2(k)(\frac{2t_1}{t_2}\cos{\frac{k}{2}})^{N_z},
\label{Eq15}
\end{align}
where $N_z$ denotes the width of zPNR. 
Therefore, the edge state wave-functions for a finite zPNR can be obtained by diagonalizing the matrix $\langle\Psi^\nu(k)|H'|\Psi^{\nu'}(k)\rangle$ for $\nu,\nu'=A,B$ which results in the following combinations of $|\Psi^A(k)\rangle$ and $|\Psi^B(k)\rangle$:
\be
|\Psi_\pm(k)\rangle =\frac{1}{\sqrt{2}}(|\Psi^A(k)\rangle \pm |\Psi^B(k)\rangle), 
\label{Psipm}
\ee 
with corresponding eigenvalues 
\be
e_\pm(k)=\pm t_2 \gamma^2(k)(\frac{2t_1}{t_2}\cos{\frac{k}{2}})^{N_z}.
\ee
We can now proceed to calculate the effect of $H''$ term in Hamiltonian(\ref{newH}) on the above energy spectrum. 
This can be done by obtaining the above matrix elements for $H''$ term in addition to $H'$. It is obvious that this only changes the diagonal elements of the matrix which gives rise to the following result for energy $E_\pm(k)$ using the $E(k)$ of Eq.(\ref{Eq4}): 
\begin{align}
&E_\pm(k)&=&E(k)+e_\pm(k) \nonumber \\
&& =& -4\frac{t_4t_1}{t_2}(1+\cos{(k)})\pm t_2 (\frac{2t_1}{t_2}\cos{\frac{k}{2}})^{N_z}\gamma^2(k).
\label{Epm}
\end{align}

\begin{figure}[t!]
\center{\includegraphics[width=1\linewidth]{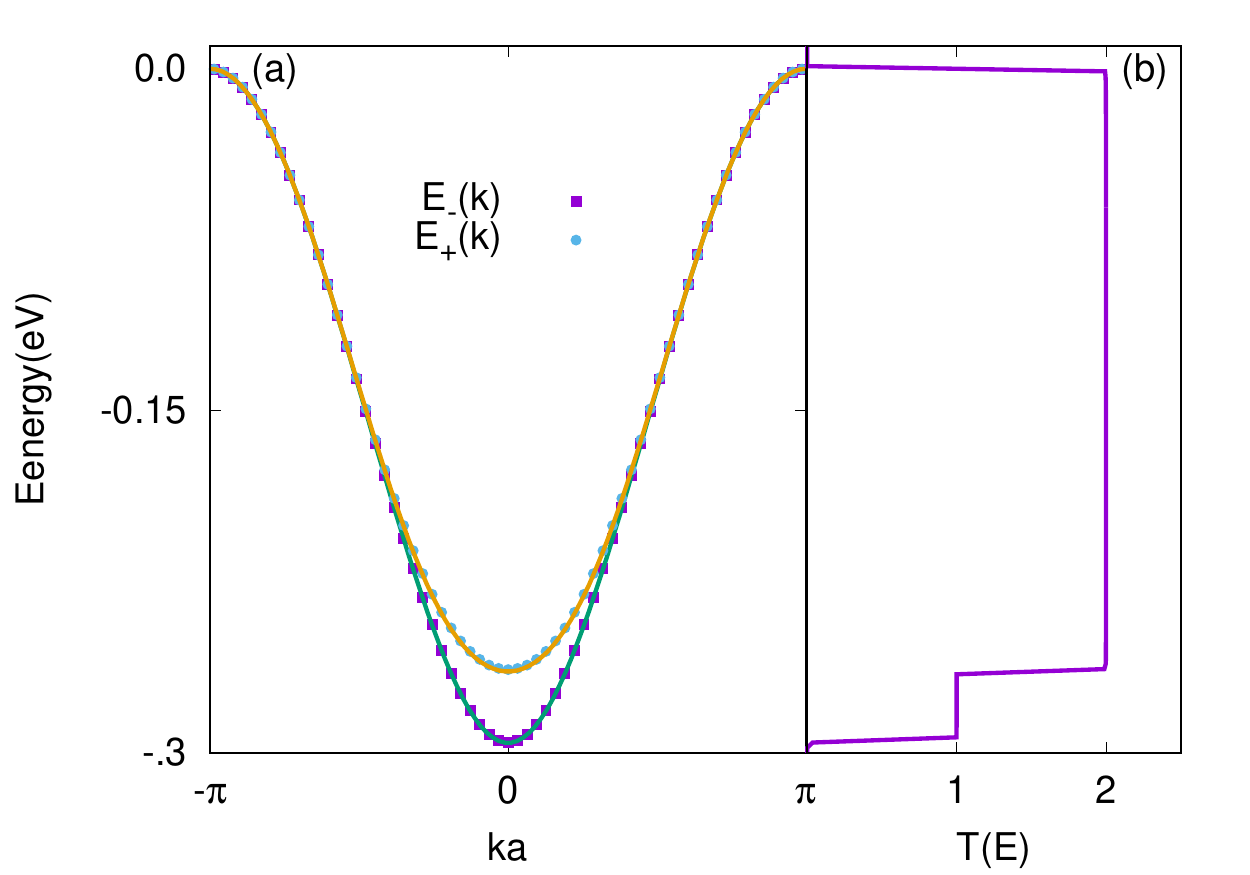}} 
\caption{(Color online) The analytical (solid lines) and numerical (points) representation of the electronic band dispersion of the edge states ($E_+(k)$ and $E_-(k)$)  for clean finite zPNR with $N_z=12$ (a) and the corresponding transmittance of the system (b).}
\label{FzPNR} 
\end{figure}

Fig.~\ref{FzPNR} (a) shows a graphical representation of the energy dispersion in Eq. (\ref{Epm}) (solid lines) for a finite zPNR with $N_z=12$ dimer chain.
It is also shown that the numerical results (points) of the edge bands which is obtained in the absence of $t_3$ and $t_5$ are fully consistent with the analytic expressions in Eq. (\ref{Epm}).
It can also be seen from the above analysis that, the corresponding wave-functions of the edge states in the finite zPNRs is a superposition of both $|\Psi^A\rangle$ and $|\Psi^B\rangle$. Therefore, the degeneracy of the edge bands will breaks~\cite{Ostahie} with a spiliting $\delta E(k)=E_+(k)-E_-(k)$ which is obvious in Fig.~\ref{FzPNR} (a).
The corresponding transmittance of the finite zPNR is presented in Fig.~\ref{FzPNR} (b).
This clearly shows a plateau of $T=2$ which denotes the presence of two transmission channel in the degenerate region for the finite zPNR. Due to the coupling of edges a slight lifting of the degeneracy takes place which causes a lower plateau of $T=1$ near the $k=0$.
It is important to note that the presence of $t_3$ and $t_5$ increases the band splitting $\delta E(k)$.
This is due to the fact that $t_3$ and $t_5$ increase the localization length of the edge states and hence increases the coupling between two edges of zPNR. 
Another significant point is that for extremely thin zPNR where the band splitting becomes of the order of
the uncoupled band widths, one of the bands changes drastically and introduces a new transmission channel\cite{Paez} which we do not consider in this study.

\subsubsection{Defective FINITE zPNR}
Let us now proceed to study the quantum transport of the finite zPNR in presence of a single point defect.
We start to evaluate the free Green's function operator $G_0$ for finite zPNR.
The matrix elements of $G_0$ are given by
\begin{align}
G_0&= \int_{-\pi}^{\pi} \frac{dk}{2\pi} \frac{|\Psi_+(k)\rangle \langle\Psi_+(k)|}{\varepsilon-\varepsilon_0-E_+(k)+i0^+}  \nonumber \\
&+\int_{-\pi}^{\pi} \frac{dk}{2\pi} \frac{|\Psi_-(k)\rangle \langle\Psi_-(k)|}{\varepsilon-\varepsilon_0-E_-(k)+i0^+},
\label{Green2}
\end{align}
where $|\Psi_\pm\rangle(k)$ and $E_\pm(k)$ are introduced in Eqs.~(\ref{Psipm}) and  (\ref{Epm}).

Similar to the case of semi-infinite zPNR 
we need to find some certain elements of the Green's function depending on the position of impurity.
Let us consider two different impurity positions: one on the outermost sites of the ribbon, $(0,0,A)$, and the other on a site which is moved one position far away from the edge on the same sub-lattice, $(0,1,A)$. 
We only discuss the calculations of the former case in this section but the same analysis can be done for the latter case easily.

Therefore, we need to find the following matrix element
\begin{align}
G_0(0,0,A;0,0,A;\varepsilon)&=\int_{-\pi}^{\pi} dk  \frac{\langle 0,0,A|\Psi_+(k)\rangle \langle\Psi_+(k)|0,0,A\rangle}{\varepsilon-\varepsilon_0-E_+(k)+i0^+} \nonumber \\
&+ \int_{-\pi}^{\pi} dk \frac{\langle 0,0,A|\Psi_-(k)\rangle \langle\Psi_-(k)|0,0,A\rangle}{\varepsilon-\varepsilon_0-E_-(k)+i0^+}. 
\end{align}
Using the definition of (\ref{Psipm}), we can rewrite this expression as
\begin{align}
G_0(0,0,A;0,0,A;\varepsilon)&=\int_{-\pi}^{\pi} dk \frac{1}{2}( \frac{\gamma^2(k)}{\varepsilon-\varepsilon_0-E_+(k)+i0^+} \nonumber \\
&+ \frac{\gamma^2(k)}{\varepsilon-\varepsilon_0-E_-(k)+i0^+}). 
\label{Int2}
\end{align}
The integration, now, can be done using the same calculation presented in the previous section which results in
\begin{align}
G_0(0,0,A;0,0,A;\varepsilon)&=\frac{\gamma^2(k_0^+)+2i\sin{(k_0^+)}}{2iE'_+(k_0^+)}\nonumber\\
&+\frac{\gamma^2(k_0^-)+2i\sin{(k_0^-)}}{2iE'_-(k_0^-)}. 
\end{align}
where $k_0^{+}$ and $k_0^{-}$ are the simple poles of the integrand in Eq. (\ref{Int2}) corresponding to $E_+(k)$ and $E_-(k)$ respectively and $E'_\pm(k)$ is the first differential of $E_\pm$ with respect to $k$.

Consider now, as before, the Lippmann-Schwinger equation (\ref{LS}) once for $|\Psi_0(k)\rangle=|\Psi_+(k)\rangle$ and once for $|\Psi_0(k)\rangle=|\Psi_-(k)\rangle$ which describes a flux incoming from the left of the defect in the channels $+$ and $-$ respectively. 
Therefore, for the $+$ channel we can write
\be
|\Psi(k)\rangle=|\Psi_+(k)\rangle+\hat{G_0}\hat{T}|\Psi_+(k)\rangle. 
\label{LS2}
\ee
We must now find
\be
\hat{G_0}\hat{T}|\Psi_+(k)\rangle= \hat{G_0} \frac{U|0,0,A\rangle\langle 0,0,A|}{1-UG_0(0,0,A;0,0,A)}|\Psi_+(k)\rangle,
\ee
which can be evaluated using the definition in Eq. (\ref{Green2}). Using again the previously mentioned method to integrate with the resulting expression, one can obtain
\begin{align}
\hat{G_0}\hat{T}|\Psi_+(k)\rangle&=\frac{1}{1-UG_0(0,0,A;0,0,A)}\frac{1}{2}(\frac{\gamma^2(k_0^+)}{E'_+(k_0^+)}|\Psi_+(k)\rangle \nonumber \\
&+\frac{\gamma(k_0^+)\gamma(k_0^-)}{E'_+(k_0^-)}|\Psi_-(k)\rangle).
\label{G_0T}
\end{align}

Substituting Eq. (\ref{G_0T}) into Eq. (\ref{LS2}), it is evident that an incident wave from the left in the $+$ channel, after the scattering, will be transmitted into right through the $+$ channel as well as the $-$ channel with the amplitudes 
\be
t_{+-}= 1+\frac{1}{2(1-UG_0(0,0,A;0,0,A))} \frac{\gamma^2(k_0^+)}{iE'(k_0^+)}
\label{t1}
\ee
and
\be
t_{+-}= \frac{1}{2(1-UG_0(0,0,A;0,0,A))} \frac{\gamma(k_0^+)\gamma(k_0^-)}{iE'(k_0^+)}
\label{t2}
\ee
respectively. 
The same analysis can be done, when the electron incident from $-$ channel and will be transmitted into $-$ and $+$ channel with amplitiudes 
\be
t_{--}= 1+\frac{1}{2(1-UG_0(0,0,A;0,0,A))} \frac{\gamma^2(k_0^-)}{iE'(k_0^-)}
\label{t3}
\ee
and
\be
t_{-+}= \frac{1}{2(1-UG_0(0,0,A;0,0,A))} \frac{\gamma(k_0^+)\gamma(k_0^-)}{iE'(k_0^-)}
\label{t4}
\ee
respectively. 
Consequently, it is now possible to introduce the transmission matrix
\be
\hat{t}=
\begin{bmatrix}
t_{++} & t_{+-} \\
t_{-+} & t_{--},
\end{bmatrix}
\label{Matt}
\ee
that allows to write the the total transmission coefficient $T$ as
\be
T=\Tr{(\hat{t}\hat{t}^\dagger)}.
\label{T_tot}
\ee

We will discuss the results of this formalism in the next section. 
\section{Results and discussion}\label{III}
In this section, we present and discuss the 
electronic transmittance through the edge states of the zPNRs in presence of a single point defect.
Using the above formalism, we also present the LDoS at the edge of ribbon near the impurity for semi-infinite zPNR.
In all the calculations of this section 
the hopping energies $t_3$ and $t_5$ considered to be zero except the last subsection.

\subsection{Results for semi-infinite zPNR}
\begin{figure}[t!]
\center{\includegraphics[width=1\linewidth]{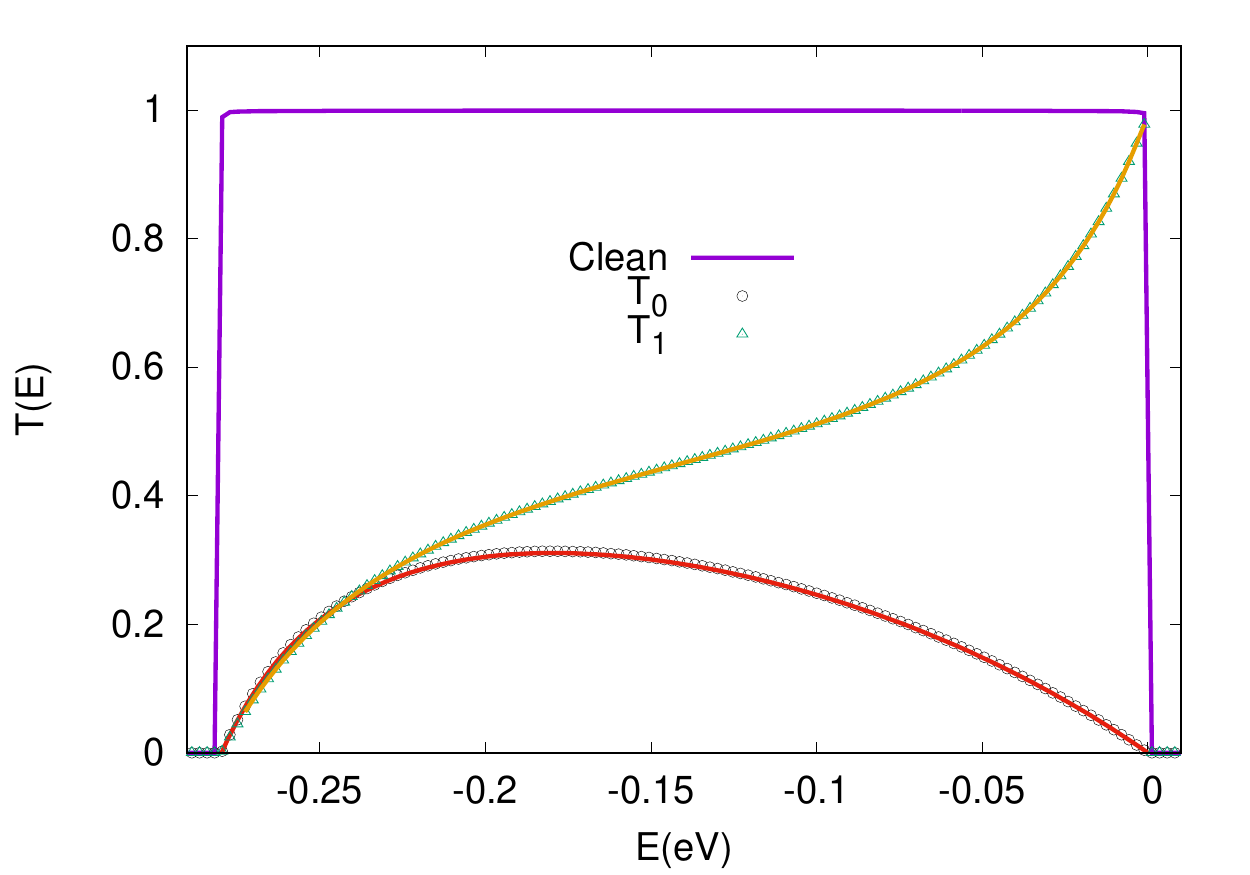}} 
\caption{(Color online) Transmittance $T_0=|\mathcal{T}_0|^2$ when the impurity which is situated on site $(0,0,A)$ and $T_1=|\mathcal{T}_1|^2$ for the impurity on on site $(0,1,A)$  both analytically (solid lines) and numerically (points) with on-site impurity potential $U=0.5(eV)$ for a semi-infinite zPNR.}
\label{T_AB}
\end{figure}

Let us first consider the case of semi-infinite zPNR with a single defect at the edge of zPNR, namely $(m=0)$. 
In general, for each $n$ (with $m=0$) there are two different sites to locate the impurity, namely $A$ and $B$, but only the outermost site ($A$ for the top edge $(m=0)$)) is relevant. This is due to the vanishing of the wave-function amplitude on the $B$ sites which we discussed before. Therefore, our first choice is the site $(n=0,m=0,A)$ to locate the impurity.
Thus, we can insert Eqs.~(\ref{G0A_D}) and (\ref{G0A_OD}) into Eq.(\ref{T_def}) which results in the following expression for the transmission amplitude:
\be
\mathcal{T}_0=1+\frac{U\frac{-i\gamma^2(k_0)}{2t'\sin{(k_0)}}}{1-U(\frac{-i\gamma^2(k_0)}{2t'\sin{(k_0)}}-\frac{(\frac{2t_1}{t_2})^2}{4t'})}.
\label{T0exp}
\ee
From now on, we use the subscript $0(1)$ of $\mathcal{T}_{0(1)}$ to show the transmission amplitiude when the impurity is located on site $(n=0,m=0(1),A)$.
Fig. (\ref{T_AB}) shows a graphical representations of the transmittance $T_0=|\mathcal{T}_0|^2$ as a function of energy (using the definition of Eq. (\ref{E_def})) for the impurity potential $U=0.5(eV)$ at the edge of the semi-infinite zPNR. 
It is also clear that the numerical results show an excelent agreement with the corresponding analytical representation of expression in Eq. (\ref{T0exp}).
We would like to emphasize that this expression coincides with the transmission coefficient of an electron through a one-dimensional channel\cite{PRB65193402} with an additional energy shift and shows the deviation of the transmission from the quantized value of a perfect channel.
Indeed, the incident electron will be totally reflected for the states near the edges ($k=0,\pm\pi$) of the quasi-flat  band and the maximum value of transmittance is achieved for the states near the band center.

There is also, however, a further position to be considered as the location of the impurity which is $(n=0,m=1,A)$ site.
Similarly, now we can insert Eqs.~(\ref{G0B_D}) and (\ref{G0B_OD}) into Eq.(\ref{LS}) to obtain
\be
\mathcal{T}_1=1+\frac{U\frac{-i\gamma^2(k_0)\alpha^2(k_0)}{2t'\sin{(k_0)}}}{1-U(\frac{-i\gamma^2(k_0)\alpha^2(k_0)}{2t'\sin{(k_0)}}
+\frac{(\frac{2t_1}{t_2})^2(1-(\frac{2t_1}{t_2})^2)}{4t'}-\frac{(\frac{2t_1}{t_2})^4\cos{(k_0)}}{8t'})}.
\ee

In Fig.~\ref{T_AB}, we present the graphical representations of the transmittance $T_1=|\mathcal{T}_1|^2$ as a function of energy for $U=0.5(eV)$. 
As it is evident there is a striking difference between
the transmittance curves $T_0$ and $T_1$. 
Namely, while the transmittance $T_1$ of
the zPNR when the impurity is located on site $(0,1,A)$ remains practically unaffected for $E\approx 0$ ($k\approx\pm\pi$), the transmittance $T_0$ when the impurity is located on the $(0,0,A)$ site, shows noticeable deviations from the clean step.

\begin{figure}[t!]
\center{\includegraphics[width=1\linewidth]{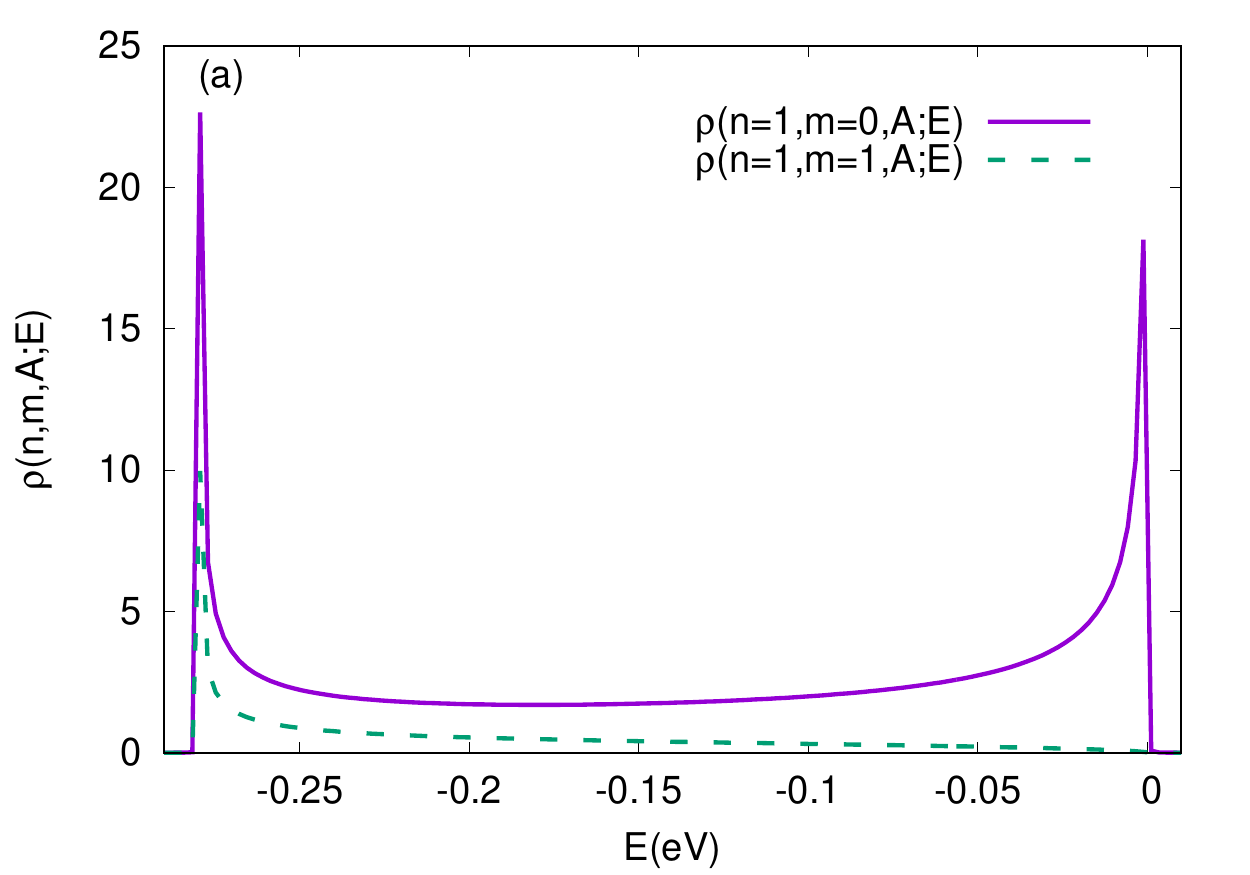}} 
\center{\includegraphics[width=1\linewidth]{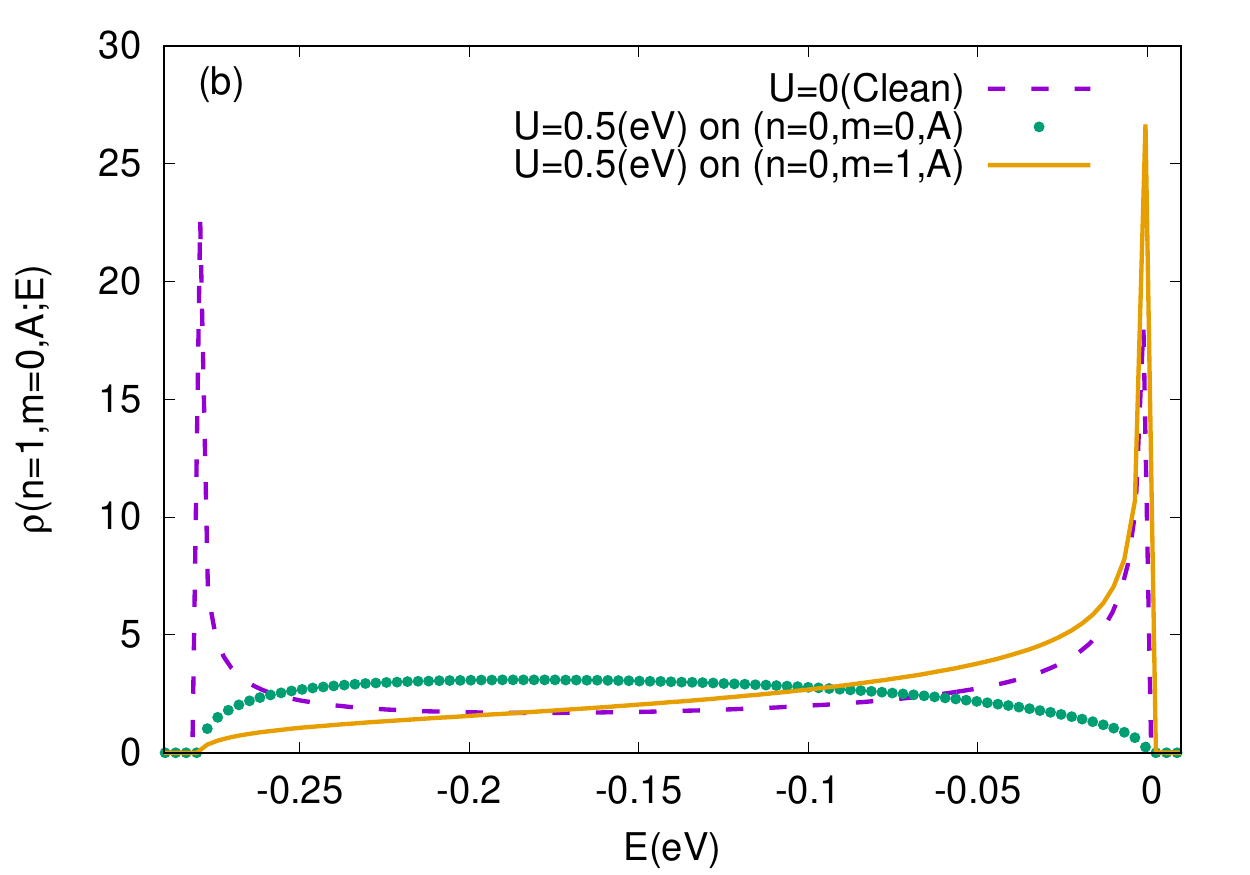}} 
\caption{(Color online) Graphical representations of the local density of states(LDoS) as a function of energy, for the semi-infinite zPNR in the absence  of defect ($U=0.0(eV)$) (a) and the same in presence of defect  ($U=0.5(eV)$) (b).}
\label{LDOS}
\end{figure}

Another significant factor in the study of transport is the behavior of the LDoS as a function of energy on the neighboring sites of the impurity position.
This is related to the imaginary part of the Green’s function through Eq.(\ref{rho-def}).
In the clean regime ($U=0$), it is straightforward to obtain the LDoS on a site at the edge of the ribbon using Eq.(\ref{G0A_OD})  which results in
\begin{align}
\rho(n=1,m=0,A;E)&=-\frac{1}{\pi} \Im{G_0(1,0,A;1,0,A;E)}\nonumber \\
&=\frac{1}{\pi} \frac{\gamma^2(k_0)}{2t'\sin{(k_0)}},
\end{align}
which is represented as a function of energy in Fig.~\ref{LDOS} (a).
It is obvious that there are two peaks at the energies close to the edges of the quasi-flat band ($k=0$ and $k\approx\pm\pi$) indicating the strong localization of corresponding states on the outermost sites of the zPNR.
As a comparison we also present the LDoS on site $(1,1,A)$ for $U=0$ which shows a strong decrease of the wave-function amplitiude for the states with momentum $k\approx \pm\pi$. 
On the other hand, the behavior of LDoS changes radically in the presence of the defect and strongly depends on the position of impurity.
We now discuss how the behavior of LDoS depends on the defect position across a nanoribbon.
When the impurity is situated on site $(n=0,m=0,A)$, by substituting
the proper Green's functions (\ref{G0A_D}), and (\ref{G0A_OD})
into Eq.(\ref{LDOS-def}), we get
\begin{align}
\rho(n=1,m=0,A;E)=\frac{1}{\pi} \frac{\gamma^2(k_0)}{2t'\sin{(k_0)}} \nonumber\\
-\frac{1}{\pi} \Im{\frac{u(\frac{-i\gamma^2(k_0)e^{ik_0n}}{2t'\sin{(k_0)}})^2}{1-u(\frac{-i\gamma^2(k_0)}{2t'\sin{(k_0)}}-\frac{(\frac{2t_1}{t_2})^2}{4t'})}}
\end{align}
which is plotted in Fig.~\ref{LDOS} (b) for $U=0.5(eV)$. 
It clearly shows the vanishing of LDOS peaks at the band edges due to the presence of impurity potential which is similar to the behavior of LDOS in a one-dimensional atomic chain with an isolated defect\cite{Datta}.
But, when the impurity is situated on site $(n=0,m=1, A)$, one position far away from the boundary of zPNR, the LDoS peak near the edge $E\approx 0$ ($k\approx\pm\pi$) remains more or less unaffected while the peak at other edges $k\approx 0$ vanishes like before. 
Here, we do not present the analytical expression explicitly for the latter case where impurity situated on site $(n=0,m=1, A)$ and we only show the graphical representations of it in Fig.~\ref{LDOS} (b).

This remarkable behavior is related to the
effective localization length of the edge state with different momentum $k$ at the zigzag edges.
According to the Eq.~(\ref{Psi_k}), the states with momentum $k=\pm\pi$ are strongly localized only on the outermost sites of the zPNR.
Therefore, an electron with the momentum $k=\pm\pi$ ($E=0$) does not scatter from the  impurity potential $U$  which is localized far away from the edges.

\subsection{Results for finite zPNR}
\begin{figure}[t!] 
\center{\includegraphics[width=1\linewidth]{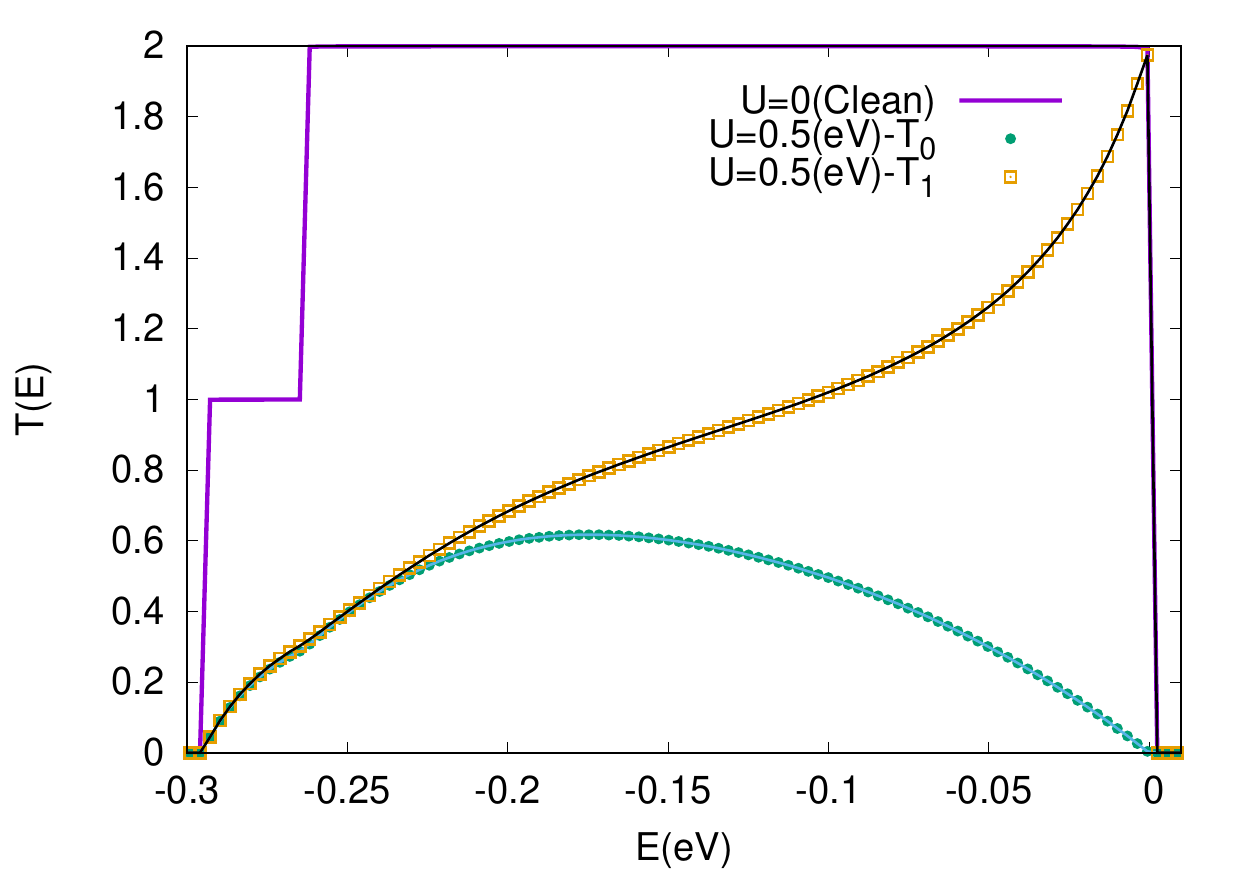}}  
\caption{(Color online) Graphical representations of  transmittance $T_0(E)$ and $T_1(E)$  both analytically (solid lines) and numerically (points) with on-site impurity potential $U=0.5$ for a finite zPNR of width $N_z=12$.}
\label{Finite1}
\end{figure}

Let us now turn to the case of finite zPNR in presence of a single defect.
Similar to the case of semi-infinite zPNR, we consider two different sites $(n=0,m=0,A)$ and $(n=0,m=1,A)$ to locate the impurity defect. When the impurity situated on site $(n=0,m=0,A)$ one can substitute Eqs. (\ref{t1}),(\ref{t2}),(\ref{t3}), and (\ref{t4}) into (\ref{Matt}) and use (\ref{T_tot}) to obtain the transmittance $T_0$ of the finite zPNR.
It is, however, straightforward to calculate the transmittance $T_1$ when the impurity position moves to site $(n=0,m=1,A)$.
The resulting expressions in the case of finite zPNR are more mathematically complicated than the semi-infinite zPNR case and we will not present here. 

Fig.~\ref{Finite1} shows the representations of resulting expressions for transmittance $T_0$ and $T_1$ for a finite zPNR of width $N_z=12$ dimer chain and scattering potential $U=0.5(eV)$. 
It is important to note  that the above-mentioned analysis holds only for the degenerate region of the band,
namely $\text{Min}(E_+(k))<E(k)<0$, and for the band splitting region which we have only one channel, $\text{Min}(E_-(k))<E(k)<\text{Min}(E_+(k))$ we need to do the same calculations as we did for one channel in the semi-infinite zPNR case.
Like the semi-infinite zPNR case, the suppression of transmittance $T_0$ and $T_1$ of an alectron with momentum $(k\approx 0)$ is observed. But, for the states near the with $kapprox\pm\pi$ ($E\approx 0$), only $T_0$ is suppressed and $T_1$ remains unaffected.

\subsection{Effect of $t_3$ and $t_5$}

\begin{figure}[t!] 
	\center{\includegraphics[width=1\linewidth]{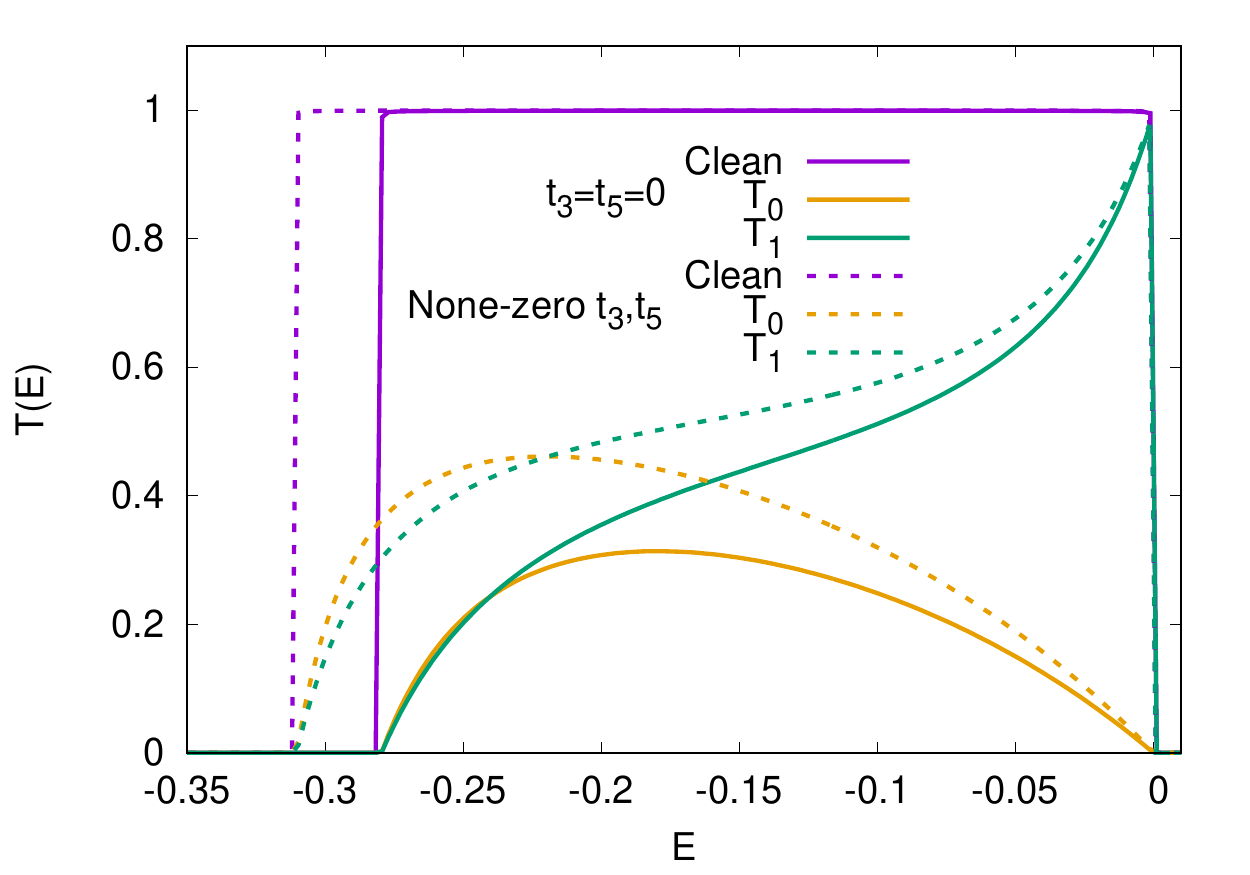}}  
	\caption{(Color online) Transmittance $T_0(E)$ and $T_1(E)$  of the semi-infinite zPNR in presence of a single impurity potential $U=0.5(eV)$ which is obtained numerically in the absence of $t_3$ and $t_5$ (solid lines) and in presence of $t_3$ and $t_5$ (dashed lines).}
	\label{nonzero}
\end{figure}

Having understood the effect of a single impurity on the transport properties of zPNRs in the absence of hopping integrals $t_3$ and $t_5$, let us now turn to the case where both $t_3$ and $t_5$ are nonzero.
For this case, we only consider the semi-infinite zPNR and perform the numerical calculation.
As before, we consider two different positions $(n=0,m=0,A)$ and $(n=0,m=1,A)$ to locate the impurity which results in the corresponding transmittance $T_0$ and $T_1$.
Fig.~\ref{nonzero} shows the effect of nonzero $t_3$ and $t_5$ on the electronic transmittance of semi-infinite zPNR both in the presence and absence of the defect.
It is obvious that there is no qualitative difference between the two regimes and the presence of $t_3$ and $t_5$ only makes the quasi-flat band wider a bit. Therefore, the corresponding values of the transmittance in this regime increases since now one needs a larger scattering potential, $U>0.5(eV)$, to suppress the transmittance as before.

\section{Conclusion} \label{IV}

In summary, we have analytically studied 
the effect of a single impurity defect on the electronic transport through the edge states of the zigzag phosphorene
nanoribbons.
Using the standard Lippmann–Schwinger T-matrix
approach, we have developed an analytical approach  
based on the tight-binding describtion of electrons on an anisotropic  honeycomb lattice. 
Our results are the following:

We approach analytically the question of quantum transport through the toplogical edge states of zPNRs which 
form a quasi-flat band in presence of a localized impurity defect.

We obtain the expressions of transmission coefficient and LDOS by calculating the lattice Green’s function for both semi-infinite and finite zPNRs.
Furthermore, we explain the features in the transmittance and obtain analytical expressions that allow us to understand the transport characteristics of zPNRs as a function of energy and scattering potential.

We present the sensitivity of elctronic transmittance through the edge states by the presence of a single point defect.
For the case where impurity is situated on the outermost sites of the ribbon, the transmission through the states with energies close to the both lower and upper edges of the quasi-flat band suppressed strongly. This is similar to behavioir of transmission in a one-dimensional chain. On the hand, when the impurity moves one position far a way from the edge, only the transmittance through the states with energies close to the lower edge of the band decreased while the states near the upper edge of the band remained unaffected.

The presented analytical expressions for transmittance and LDOS are in good agreement with numerical calculations based on Landauer
approach. 

We also perform numerical computations of transmittance using the Landauer approach which showes a very good agreemen with analytical expressions. We further compute the effect of hopping integrals $t_3$ and $t_5$ which show no signicant effect on the transport through the edges in both clean and defective samples.

Finally, we should emphasize that in this paper, we focused on the effects of a single point defect with finite scattering potential. 
However, it is also intersting to study the effect of vacancy as the limiting case $U\rightarrow\infty$ on the transport properties of the phosphorene ribbons which induces compact localized states with intersting impacts on the transport.
These results which are beyond the aim of this paper will be discussed in more detail elsewhere.



{\it Acknowledgment} 
We gratefully acknowledge R. Asgari and E. Ghanbari-Adivi for useful discussions and comments during improvement of this work. We also would thank the International Center for Theoretical Physics
(ICTP), Trieste, Italy for their hospitality and support during a visit in which part of this work was done.

\end{document}